
\input phyzzx.tex

\def\prdj#1{{\it Phys. Rev.} {\bf D{#1}}}
\def\npbj#1{{\it Nucl. Phys.} {\bf B{#1}}}
\def\prlj#1{{\it Phys. Rev. Lett.} {\bf {#1}}}
\def\plbj#1{{\it Phys. Lett.} {\bf B{#1}}}
\def\zpcj#1{{\it Z. Phys.} {\bf C{#1}}}

\def\prepj#1{{\it Phys. Rep.} {\bf {#1}}}

\def\anti{\overline}
\def\pbi{~{\rm pb}^{-1}}
\def\fbi{~{\rm fb}^{-1}}

\catcode`\@=11 

\def\t1{{\tilde 1}}
\def\ov{\overline}

\def\slash#1{#1\hskip-6pt/\hskip6pt}
\def\ptmiss{\slash p_T}

\def\gev{\,{\rm GeV}}
\def\tev{\,{\rm TeV}}

\def\wt{\widetilde}

\def\rta{\rightarrow}
\def\mhalf{m_{1/2}}
\def\gl{\wt g}
\def\mgl{m_{\gl}}
\def\stop{\wt t}
\def\mstop{m_{\stop}}
\def\sq{\wt q}

\def\msq{m_{\sq}}
\def\slep{\wt \ell}

\def\mslep{m_{\slep}}

\def\hl{h^0}
\def\hh{H^0}
\def\ha{A^0}
\def\hp{H^+}
\def\hm{H^-}
\def\hpm{H^{\pm}}
\def\mhl{m_{\hl}}
\def\mhh{m_{\hh}}
\def\mha{m_{\ha}}

\def\tanb{\tan\beta}
\def\mt{m_t}
\def\mb{m_b}
\def\mz{m_Z}
\def\mgut{M_U}

\def\wp{W^+}
\def\wm{W^-}

\def\wmp{W^{\mp}}
\def\cnone{\wt\chi^0_1}
\def\cntwo{\wt\chi^0_2}
\def\snu{\wt\nu}
\def\snubar{\ov{\snu}}
\def\msnu{m_{\snu}}
\def\mcnone{m_{\cnone}}

\def\h{h}
\def\cpone{\wt \chi^+_1}
\def\cmone{\wt \chi^-_1}
\def\cpmone{\wt \chi^{\pm}_1}
\def\mcpone{m_{\cpone}}
\def\stauone{\wt \tau_1}
\def\mstauone{m_{\stauone}}

\date{January, 1994}
\Pubnum{$\caps UCD-94-1$\cr}
\titlepage
\baselineskip 0pt
\hsize=6.5in
\vsize=8.5in
\bigskip
\centerline{\bf EXPLORING THE YUKAWA UNIFIED MINIMAL}
\vskip .05in
\centerline{\bf SUPERGRAVITY MODEL AT THE TEVATRON, LEP II AND THE LHC}
\smallskip
\centerline{J.~F.~GUNION and H. POIS}
\smallskip
\centerline{\it Davis Institute for High Energy Physics,}
\centerline{\it Department of Physics, U. C. Davis, Davis, CA 95616}
\vskip .075in
\centerline{\bf Abstract}
\vskip .075in
\centerline{\Tenpoint\baselineskip=12pt
\vbox{\hsize=12.4cm
\noindent We explore the prospects for detection of sparticles and Higgs bosons
at the Tevatron, LEP-200 and the LHC in the  allowed parameter space of
a `yukawa unified' ($\lambda_b(M_U)=\lambda_{\tau}(M_U)$)
minimal supergravity (YUMS) model, where the only non-zero unification scale
soft-SUSY-breaking terms are a universal gaugino
mass and a Higgs mixing term. In a bottom-up approach, just two weak scale
parameters, $\tanb$ and $\mha$, (along with the sign of the Higgs mixing
parameter $\mu$) completely parameterize the model. Many interesting
`special' situations regarding sparticle and Higgs discovery arise,
such as the importance of the invisible $\hl\rta\cnone\cnone,\snu\snubar$
decay modes.
}}

\vskip .15in
\leftline{\bf 1. Introduction}
\vskip .075in
Incorporation of the Standard Model (SM) into a more fundamental
theory that includes supersymmetry (SUSY) is both aesthetically
attractive and a theoretically compelling solution to
the naturalness and hierarchy problems.
Although sparticles have not been directly detected to date,
and we still lack a realistic mechanism for SUSY breaking,
the success of SUSY gauge coupling unification and
the accurate prediction of $\sin\theta_W$,
\Ref\unif{J. Ellis, S. Kelley and D. V. Nanopoulos, \plbj{249} (1990) 441;
U. Amaldi, W. de Boer and A. Furstenau, \plbj{260}(1991) 447;
P. Langacker and M.-X. Luo, \prdj{44} (1991) 817.}\
along with an elegant radiative electroweak symmetry breaking (EWSB)
scenario\Ref\NSrep{Lahanas and D. V. Nanopoulos, \prepj{145} (1987) 1.}\
(which requires a heavy top quark)
all provide dramatic indirect support for SUSY.

Squark, gluino, slepton, chargino and
neutralino production, decay and detection have been studied in
a generic fashion at LEP-200, Fermilab and hadron
supercolliders,\Ref\sparticles{We give only a few representative surveys;
further references can be found in the papers cited.
T. Tsukamoto \etal, KEK-PREPRINT-93-146 (1993);
J.F. Grivaz, preprint LAL-93-11;
M.M.J.F. Janssen, \zpcj{53} (1992) 453;
A. Amos \etal, Proceedings
of {\it Research directions for the Decade}, Snowmass (1990), p. 516;
H. Baer \etal, \ibid\ p. 182; V. Fonseca \etal, \ibid\ p. 198;
R.M. Barnett \etal, \ibid\ p. 201; J.F. Gunion and H. Haber, \ibid\ p. 206;
A. Bartl \etal, Proceedings of {\it The Large Hadron Collider Workshop},
p. 615; C. Albajar \etal, \ibid\ p. 621; H. Baer \etal, \ibid\ p. 654;
R. Barbieri \etal, \ibid\ p. 658; F. Del Aguila \etal, \ibid\ p. 663.}\
\REF\flipped{R. Arnowitt and P. Nath, preprint CTP-TAMU-32-93;
J.L. Lopez, D.V. Nanopoulos, G.T. Park, X. Wang, and A. Zichichi,
preprint CERN-LAA-94-01.}\
as well as in supergravity GUT models at LEP-200,
Fermilab and HERA\refmark{\flipped}.

The phenomenology of the five physical Higgs bosons (the $\hl,\hh,\ha,\hpm$)
of the Minimal Supersymmetric Model (MSSM),\Ref\Hunters{For a survey and
references, see J. F. Gunion, H. E. Haber, G. L. Kane and
S. Dawson, {\it The Higgs Hunter's Guide}, Addison-Wesley,
Redwood City, CA (1990).}\ has also been examined at
$\epem$\Ref\epemrev{For a review and references, see J.F. Gunion,
preprint UCD-93-24, to appear in {\it Proceedings of the
2nd International Workshop on ``Physics and Experiments with
Linear Colliders"}, ed. F. Harris, Waikoloa, HI, April 26-30 (1993).}
and hadron\Ref\SUSYhiggs{H. Baer \etal, \prdj{46}
(1992) 1067; J.F. Gunion and L. Orr, \prdj{46} (1992) 2052; V. Barger,
K. Cheung, R.J.N. Phillips and A. Stange, \prdj{46} (1992) 4914; Z. Kunszt and
F. Zwirner, \npbj{385} (1992) 3.}\
\REF\radcorra{H. Haber and R. Hempfling \prlj{66} (1991) 1815; Y. Okada, M.
Yamaguchi and T. Yanagida, {\it Prog. Theor. Phys.} {\bf 85} (1991) 1;
J. Ellis, G. Ridolfi and F. Zwirner, \plbj{257} (1991) 83.}
supercolliders, including one-loop\refmark{\radcorra}\ radiative corrections.
\foot{Typically evaluated with
$\mstop=1\tev$, for which two-loop radiative corrections may also be
important.\Ref\radcorrb{R. Hempfling and A. Hoang,
preprint DESY-93-162 and TTP93-35.}}
\REF\jfgperspectives{J.F. Gunion, `Detecting the SUSY Higgs Bosons', {\it
Perspectives in Higgs Physics}, ed. G. Kane, World Scientific Publishing
(1992), p. 179.}
\REF\JFGproc{J. F. Gunion and S. Geer, `Progress in SSC Higgs Physics:
Report of the Higgs Working Group', to appear in Proceeding of the "Workshop
on Physics at Current Accelerators and the SuperCollider", eds. J. Hewett,
A. White, and D. Zeppenfeld, Argonne National Laboratory, 2-5 June (1993).}
For recent reviews of the hadron collider phenomenology, and further
references, see Refs.~[\jfgperspectives,\JFGproc].
If we assume that SM final state modes dominate Higgs decays, the discovery
channels of most obvious utility at hadron colliders are:
i) $\hl,\hh,\ha\rta \gamma \gamma$;
ii) $l(\hl,\hh\rta \gamma \gamma)$;
iii) $\hl,\hh\rta 4l$;
iv) $t\rta \hp b$; and possibly
v) $\hh,\ha\rta \tau^+\tau^-$.
\REF\JFGhplus{J.F. Gunion, preprint UCD-93-40 (1993).}
\REF\JFGbbbar{J. Dai, J.F. Gunion, and R. Vega, \prlj{71} (1993) 2699.}
Further, $\sim 30\%$ efficiency and $\sim 1\%$ purity for $b$-tagging will
allow neutral and charged Higgs detection in
the production/decay modes $t\bar t (\hl,\hh\rta b\bar b)$\refmark\JFGbbbar\
and  $t \bar b (\hp\rta t\bar b)$\refmark\JFGhplus\ for some parameter choices.
In particular, the $t\anti t(\hl,\hh\rta b\anti b)$
mode combined with i)-iv) above implies
\Ref\nolose{J. Dai, J. F. Gunion and R. Vega, \plbj{315} (1993) 355.}\
that there is a
no-lose theorem, according to which detection of at least one of the MSSM Higgs
bosons will be possible at the LHC. However, over much of
parameter space, in particular for $\mha\gsim 100-200\gev$, only the $\hl$ will
be easily detected, and it will be relatively SM-like.

\REF\ghiii{J.F. Gunion and H.E. Haber, \npbj{307} (1988) 445 (E: \npbj{402}
(1993) 569). See also Ref.~[\Hunters].}
\REF\susyinvisible{K. Greist and H.E. Haber, \prdj{37} (1988) 719.}
\REF\baersparticle{H. Baer, M. Bisset, D. Dicus,
C. Kao, and X. Tata, \prdj{47} (1993) 1062.}
\REF\jfgsparticle{J.F. Gunion, `Detecting the SUSY Higgs Bosons',
Proceedings of the 23rd INFN Eloisatron Workshop on {\it Properties of
SUSY Particles}, eds. L. Cifarelli and V. Khoze, World Scientific
Publishing, preprint UCD-93-8.}
\REF\djouadietal{A. Djouadi, J. Kalinowski and P.M. Zerwas,
\zpcj{57} (1993) 569.}
However, if the branching ratios for $\hl,\hh,\ha,\hpm \rta {\rm sparticles}$
are large, Higgs detection in direct SM final state decay modes will become
more
difficult, while new opportunities in the sparticle
channels will arise.\refmark{\ghiii-\djouadietal}
Supergravity theories
\Ref\kaneetal{For a recent treatment, with references to all previous work, see
G. L. Kane, C. Kolda, L. Roszkowski and J. Wells, UM-TH-93-24.}\
provide a very attractive framework for evaluating
the effects of sparticle pair channels, while simultaneously
allowing an analysis
of sparticle decays as they affect not only the detection of such indirectly
produced sparticles but also direct sparticle searches.
Assuming coupling constant unification and
universal soft-SUSY breaking at the unification scale $\mgut$, just {\it five}
independent parameters are needed to specify the models.
One possible set comprises
$\lambda_t(\mgut),\mhalf,A,m_0,\mu_0=\mu(\mgut)$.
\Ref\elliszwirner{See, \eg, J. Ellis and F. Zwirner, \npbj{338} (1990) 317.}\
Here, $m_0,A,\mhalf$ are the (common) scalar, $A$-term and gaugino masses at
$\mgut$, respectively. Alternatively, a bottom-up approach can be employed
in which all (five) input parameters are specified at the weak scale.
\Ref\pokorski{M. Olechowski and S. Pokorski, \npbj{404} (1993) 590.}\

In this work, we adopt a bottom-up approach but simplify the parameter space
further by exploring the `yukawa unified'
($\lambda_b(\mgut)=\lambda_\tau(\mgut)$) minimal supergravity (YUMS) model
in which the soft-SUSY-breaking parameters $m_0$ and $A$ are taken
to be zero. The particle content is that of the MSSM:
SM fields; two Higgs doublets; and their superpartners.
We require gauge coupling unification, but
do not insist on any particular GUT embedding (thus, proton decay
need not be a difficulty). The combination of
gauge and yukawa unification implies that a choice for
$\tanb$ determines a value for $\mt$,
independently of the soft-SUSY breaking parameters.
\REF\bargera{V. Barger, M. Berger, P. Ohmann and R.J.N. Phillips,
\plbj{314} (1993) 351.}
(Higgs phenomenology could be pursued at this point, for generic choices
of the remaining parameters; see \eg\ Ref.~[\bargera].)
If in addition we require $m_0=A=0$, specification of
just $\tanb$ and $\mha$ (at low-energy), along with the sign of $\mu$,
completely determines the supergravity model.
Although we allow for $B_0\equiv B(\mgut)\neq0$, the requirement
$B_0=0$ yields an interesting
special case of the YUMS model.  (Indeed, if one wants
to assume for simplicity that only one type of soft SUSY breaking is dominant,
the only possibility not ruled out by theoretical and/or phenomenological
constraints is to have $\mhalf\gg m_0,A,B_0$.)
In this case, specifying $\tanb$ alone (and the sign of $\mu$)
is sufficient to parameterize the model, since there
is a unique value for $\mha$ such that $B_0=0$.
Full details regarding our bottom-up approach to the renormalization
group equations will be presented in a future paper.
\Ref\gplater{J.F. Gunion and H. Pois, in preparation.}\

The YUMS boundary conditions have significant motivation.
Yukawa unification is especially motivated in that:
i) the constraint arises naturally in GUT models
where the $b,\tau$ fermions both reside in the same
representations; and ii) the $\lambda_b(M_U)=\lambda_{\tau}(M_U)$ condition
constrains the model to be close to the RGE-fixed point region
where $\lambda_t(M_U)\gsim 1$,\Ref\bardeennew{W.A. Bardeen,
M. Carena, S. Pokorski and C.E.M. Wagner, preprint MPI-Ph-93-58.}\
and naturally predicts a large top quark mass.\foot{Note, however, that
unification of all three yukawas, $\lambda_{b,\tau,t}$, is not
possible in the YUMS context; in particular,
$\lambda_b(M_U)\gsim 1$ requires $\tanb$ values larger than allowed.}

\REF\kapl{V. Kaplunovsky and J. Louis, \plbj{306} (1993) 269.}
As for the $m_0=A=0$ assumption, Ref.~[\kapl] notes that soft mass and
tri-linear terms are generally not flavor or generation universal
in generic supergravity models,
and therefore may lead to large flavor-changing neutral
currents.\Ref\iblu{L. Ibanez and D. Lust, \npbj{382} (1992) 305.}\
If the scalar masses are required to be {\it zero} at $\mgut$,
then the RGE's automatically maintain sufficient
degeneracy among the squarks to suppress FCNC's. In the context
of supergravity, the $m_0=A=0$ boundary condition arises naturally
in the `no-scale' theories.\refmark\NSrep\ However, we must note that
for the {\it known}
explicit realizations of this boundary condition at the string scale
in superstring theories,\foot{See, for example, the moduli-dominated
($\theta=0$) limit of Eq.~(6.4)
\REF\ibanez{A. Brignole, L.E. Ibanez and C. Munoz, preprint FTUAM-26/93.}
in Ref.~[\ibanez]; for the case where all modular weights $n_i=-1$
one finds $m_0=A=0$ at the string scale.}
threshold corrections are apparently not large
enough to allow gauge coupling unification and $m_0=A=0$
at the lower $\mgut$ scale. However, employing these boundary conditions
at $\mgut$ could still be a reasonable first approximation.

\REF\LNPWZ{J. L. Lopez, D. V. Nanopoulos, H. Pois, X. Wang and A. Zichichi,
\prdj{48} (1993) 4062.}
Given the boundary condition $m_0=0$, comparatively low masses
(\eg\ $m_{\tilde l_R}\simeq 0.15-0.19 \mgl$, and $\msq\simeq 0.86-0.92 \mgl$)
are predicted for the $\tilde l,\tilde q$ SUSY partners. This has crucial
implications for Higgs/sparticle phenomenology.
As an important example, in addition to the invisible decay
$\hl\rta \cnone \cnone$
(discussed, for instance, in Refs.~[\susyinvisible,\djouadietal,\LNPWZ])
we find that the (invisible) $\hl\rta\snu\snubar$ decay
is also kinematically accessible in a portion of the parameter
space and can even dominate over the $\hl\rta b \anti b$ mode.
These, and other sparticle decay modes of the Higgs bosons, must
be included in assessing Higgs detection in the context of $m_0\sim 0$ models.

We note that the possibility of detecting a generic
invisibly-decaying Higgs boson at a hadron supercollider, \ie\
$\h\rta I$ decay (where $I$ is an
unspecified  invisible channel), has been explored.
The two possible detection modes at a hadron supercollider
are $t\bar t+\h\rta \ell+jets+\ptmiss$\Ref\JFGinvis{J.F. Gunion,
\prlj{72} (1994) 199.}\
and $W+\h\rta \ell+\ptmiss$,\Ref\kaneroy{S.G. Frederiksen, N.P. Johnson,
G.L. Kane, and J.H. Reid, preprint SSCL-577-mc (1992);
D. Choudhury and  D.P. Roy, preprint TIFR-TH-93-64 (1993).}\
with the optimistic conclusion that even an invisibly decaying Higgs can be
detected so long as its $\h t\bar t$ and $\h \wp\wm$ couplings, respectively,
are not too suppressed compared to SM strength. Variation of these
couplings can be determined in our constrained YUMS model, and
the invisible mode procedures of Ref.~[\JFGinvis] are incorporated
in our analysis.

\FIG\one{
We display the allowed region of $\tanb,\mha$ for the YUMS
model.  The right-hand vertical axis gives the $\mt(\mt)$ value corresponding
to a given $\tanb$. Within this region we show: (short dashes) the contour for
$\mcpone=100\gev$; (solid) to the right of the contour
fewer than 100 $Z\hl$ events occur at LEP-II
(with $L=500\pbi$); (dotdash) to the right of the contour LEP-II yields
fewer than 100 $\stauone^+\stauone^-$
events --- also $\mcpone>165\gev$ to the right of this same
contour; (dots) to the right of the contour $\mgl>300\gev$;
(dotdotdash) to the left of the contour
$BR(\hl\rta\cnone\cnone+\snu\snubar)>0.2$;
(long dashes) the contour above which $\cpone\rta \ell^+\snu$
is allowed and the only allowed two-body
$\snu$ decay is the invisible $\snu\rta \nu \cnone$ (below, $\snu\rta
\ell^{\mp}\cpmone$ is allowed); (long dashes, triple dots) the contour
below which $\Delta\rho/\rho<0.01$. We have taken $\mu>0$ for this figure.}
\topinsert
\vbox{\phantom{0}\vskip 5.0in
\phantom{0}
\vskip .5in
\hskip -20pt
\special{ insert user$1:[jfgucd.pois]noscale_lep.ps}
\vskip -1.45in }
\centerline{\vbox{\hsize=12.4cm
\Tenpoint
\baselineskip=12pt
\noindent
Figure~\one:
We display the allowed region of $\tanb,\mha$ for the YUMS
model.  The right-hand vertical axis gives the $\mt(\mt)$
value corresponding
to a given $\tanb$. Within this region we show: (short dashes) the contour for
$\mcpone=100\gev$; (solid) to the right of the contour
fewer than 100 $Z\hl$ events occur at LEP-II
(with $L=500\pbi$); (dotdash) to the right of the contour LEP-II yields
fewer than 100 $\stauone^+\stauone^-$
events --- also $\mcpone>165\gev$ to the right of this same
contour; (dots) to the right of the contour $\mgl>300\gev$;
(dotdotdash) to the left of the contour
$BR(\hl\rta\cnone\cnone+\snu\snubar)>0.2$;
(long dashes) the contour above which $\cpone\rta \ell^+\snu$
is allowed and the only allowed two-body
$\snu$ decay is the invisible $\snu\rta \nu \cnone$ (below, $\snu\rta
\ell^{\mp}\cpmone$ is allowed); (long dashes, triple dots) the contour
below which $\Delta\rho/\rho<0.01$. We have taken $\mu>0$ for this figure.
}}
\endinsert

\vskip .15in
\leftline{\bf 2. Phenomenology of the $\lambda_b(M_U)=\lambda_\tau(M_U)$
Minimal Supergravity Model}
\vskip .075in

In this section, we present the basic sparticle and Higgs boson
phenomenology of the YUMS model at the Tevatron, LEP-II and the LHC.
\foot{Related work on the phenomenology of models with the $m_0=A=0$
boundary conditions in the context of no-scale models can be found in
\REF\NSphen{S. Kelley, J. L. Lopez,
D. V. Nanopoulos, H. Pois and K. Yuan \plbj{273} (1991) 423;
S. Kelley, J. Lopez, D. V. Nanopoulos,
H. Pois, K. Yuan, \npbj{303} (1993) 3; V. Barger, M. S. Berger, and
P. Ohmann, MAD/PH/801; D. J. Castano, E. J. Piard and P. Ramond,
UFIFT-HEP-93-18.}\
Refs.~[\NSrep,\NSphen].}
We begin by determining the region of $\tanb,\mha$ parameter space
that is allowed by current experimental limits on the masses of
the various sparticles and Higgs bosons in combination with the requirements of
a neutral LSP, correct symmetry breaking
(including the demand that $V<0$ at the
minimum) and a bounded Higgs potential at $\mgut$.
A precise delineation of the input limits and
experimental constraints will be given in Ref.~[\gplater].
\REF\dmnew{S. Kelley, J.L. Lopez, D.V. Nanopoulos, H. Pois, and K. Yuan,
\prdj{47} (1993) 2461.}
Considerations of dark matter/cosmology
are not included in our analysis; however, we do not expect this to be at all
limiting.\refmark\dmnew\
Constraints from $BR(b\rta s\gamma)$ do not further restrict
our parameter space if full account is taken of both the experimental
errors and the significant uncertainties in the theoretical computation
\Ref\buras{A.J. Buras, M. Misiak, M. Munz and S. Pokorski,
preprint MPI-Ph/93-77.}\ (mainly
arising from strong dependence on the undetermined
renormalization scale $\mu_{QCD}$). Finally, we note that
the proton lifetime can be made acceptably large in the YUMS model.
For example, non-minimal $SU(5)$ models\Ref\rossnm{See, for example,
G. Ross, {\it Grand Unified Theories}, Frontiers in Physics,
Benjamin/Cummings (1985).} can be constructed in which
dangerous dimension-5 graphs do not arise. However, minimal $SU(5)$
embedding would be problematical because of the large super-heavy
triplet masses required and the resulting mass hierarchy problems.
For our figures, we present the $\mu>0$ case.
The allowed region of yukawa-unified minimal supergravity (YUMS)
parameter space is displayed in Fig.~\one. The
left and right sides of the solid line boundary are essentially obtained by the
requirements $\mslep>45\gev$  and
$\mstauone>\mcnone$ (\ie\ a neutral LSP), respectively. The bottom edge of the
allowed region is a result of the LEP-I Higgs constraint $\mhl\gsim 60\gev$.
We note that for $\tanb\simeq 1.5$ our $\mha$ limits agree with the
third reference in Ref.~[\NSphen].

Note that the combination of the yukawa unification and $m_0=A=0$ boundary
conditions at $\mgut$
and the general phenomenological constraints has severely
limited the range of $\tanb,\mha$ values allowed.  The precise region
is somewhat sensitive to the values taken for $m_b(m_b)$ and $\alpha_s(\mz)$.
However, for our choices of $m_b(m_b)=4.25\gev$
\Ref\mbref{J. Gasser and H. Leutwyler, \prepj{87}
(1982) 77, find $\mb=4.25\pm 0.10\gev$.}\
and $\alpha_s(\mz)=0.12$, we find $1.5\lsim\tanb\lsim 10$,
and $185 \gev\lsim \mha \lsim 690 \gev$.  Due to the yukawa unification
condition, this immediately implies that $\mt\gsim 155\gev$.
As $m_b(m_b)$ is increased, or $\alpha_s(\mz)$ is decreased,
slightly higher $\tanb$ values are allowed. Although not shown,
for the $\mu<0$ case the right boundary shifts considerably
to the left since $\mcnone$ increases in the $\mu<0$ case. This makes the
neutral LSP restriction even more severe so that $\mha$ must be $\lsim
380\gev$.
Finally, adding the $B_0=0$ constraint to the YUMS
boundary conditions yields a
relation between $\tanb$ and $\mha$ (see Fig.~2) which is very much within the
allowed parameter space.

Limitations on the parameter space from
$\rho\equiv\mw^2/(\mz^2\cos^2\theta_W)$ can also be significant
in yukawa-unified approaches that lead to parameter
space regions with large $\mt$ values. In the YUMS model, our particular
choices of $\mb(\mb)=4.25\gev$ and $\alpha_s(\mz)=0.12$ imply that part
of the otherwise-allowed parameter space corresponds to $\mt$ values
sufficiently
large as to be disfavored by $\Delta\rho$. In Fig.~\one, we show the contour
for $\Delta\rho/\rho=0.01$. At large $\mha$, contributions to $\Delta \rho$
are dominated by the top/bottom loop and $\mt\lsim 180\gev$ ($\tanb\lsim 3$)
is favored (as in the SM). As $\mha$ becomes smaller, stop/sbottom
loop corrections also enter and the favored upper limit on $\mt$
(and $\tanb$) decreases. However, if we adopt $\alpha_s(\mz)=0.11$,
the maximum $\mt$ allowed by all other constraints is $\lsim 180\gev$,
and no additional restriction on allowed parameter
space would result from requiring $\Delta\rho/\rho\lsim 0.01$.

A significant feature of the YUMS model, which sets it apart
from models with $m_0\gsim \mhalf$, is that squark masses are driven
by the RGE's to much higher values than are the slepton masses ---
all start from $\sim 0$ at $\mgut$, but only the squark masses have the strong
$SU(3)$ group $g_s^2\mgl^2$ driving terms.
The result, is a mass ordering roughly described by
$\mcnone\simeq 0.08-0.14 \mgl$,  $m_{\tilde l_R}\simeq 0.15-0.19 \mgl$,
$\mcpone\simeq 0.16-0.26\mgl$, and $\msq\simeq 0.86-0.92 \mgl$.
This means that the decays of sparticles and Higgs bosons will be
dominated by unexpected channels.
For instance, since charged slepton masses are almost always lower
than the mass of the lightest
chargino,\foot{The only exception occurs for $\tanb\lsim 3.1$, for $\mha$
values immediately adjacent to
the left hand border of the allowed region in Fig.~\one.}
the two-body decay $\cpone\rta \slep^+ \nu$ will almost
always be present and will dominate over the three-body (mainly hadronic)
modes.
Over most of the parameter space the sneutrino is also lighter than
the chargino; the relevant contour is shown in Fig.~\one.
This implies that $\cpone\rta \ell^+\snu$ decays will also be
important. In this same region, the sneutrino
decays entirely invisibly ($\snu\rta\cnone\nu$). Note that throughout
the allowed parameter space,
the second neutralino, $\cntwo$, with mass similar to that of the $\cpone$,
has {\it both} $\snu \nu$ and $\slep \ell$ decays, whereas
$\cnone\hl$ and $\cpmone \wmp$ are never two-body allowed.
Meanwhile, the sleptons
almost always decay via $\slep^+\rta \ell^+ \cnone$, the only
exception being the above-noted border region where $\slep^+\rta\nu\cpone$
is allowed (but with small phase space). Thus, $\slep\,$'s are easily visible.
Further, a decent fraction of $\cntwo$ decays are to $\slep\ell$, which
thus yields {\it several} reasonably energetic (\ie\ visible)
charged leptons.

Squarks, being lighter than the gluino, will play a prominent role,
and will decay primarily via $\sq\rta q\wt\chi$, where $\wt\chi=\cpone$
has a large share of the branching ratio and subsequently
decays as described above.  Gluino searches will focus
on the two-body channel $\gl\rta q \sq$, followed by $\sq$ decay.

Higgs decays also have unexpected features.
For $\mha\lsim 250\gev$, the invisible channels
$\hl\rta \cnone\cnone,\snu\snubar$ can be
significant, depending on the value of $\tanb$.
Fig.~\one\ shows the region where $BR(\hl\rta \cnone\cnone+\snu\snubar)>0.2$.
Indeed, for $\tanb\gsim 6$ and $\mha\lsim 200\gev$, we find that
$BR(\hl\rta\snu\snubar)\gsim 0.9$ so that invisible modes
overwhelm all other decay modes.
\foot{In the small-$\mha$, large-$\tanb$
corner of parameter space the $\stauone^+\stauone^-$
mode can also occur, but does not have large branching ratio.}
(However, the only allowed invisible mode in the $\Delta\rho/\rho<0.01$
region is $\hl\rta\cnone\cnone$.)
Sparticle-pair channels, $\wt\chi\wt\chi$ and $\slep\slep$,
are even more important for the $\hh,\ha,\hpm$
Higgs bosons. (Squark pair modes are not present in the allowed parameter
space region due to the relatively large $\msq$ values.)
For the $\hh$, the visible $\wt\chi\wt\chi$ modes
have a branching ratio of order 40\% for much of parameter space,
while the visible $\slep\slep$ modes are generally below 5\%.
For the $\ha$, visible $\slep\slep$ modes are always insignificant
since they are determined by mixing proportional to fermion Yukawa couplings,
but visible $\wt\chi\wt\chi$ channel
branching ratios between 80\% and 90\% are quite common.
In contrast, the branching ratio for
$\hpm$ decays to (visible) $\slep^{\pm}\snu$ modes is commonly in the
50\% to 90\% range, while visible $\wt\chi\wt\chi$ channels seldom exceed 5\%.
More details will appear in Ref.~[\gplater].

At LEP-II, the most promising sparticle searches will involve
$\cpone\cmone$ and $\slep^+\slep^-$ pair production. In Fig.~\one\
we show the contour for $\mcpone=100\gev$,
the absolute upper limit for $\cpmone$
detection for a LEP-II energy of $200\gev$.
The fact that $\msnu<\mcpone$ over most of the allowed parameter space (see
the contour in Fig.~\one) implies that
the $\epem\rta\cpone\cmone$ production
rate will be lowered due to the destructive interference of the $\snu$
exchange diagram. However, the $\cpone$ decays are (as noted above)
two-body and yield energetic leptons. This means $\cpone\cmone$ pair
production will be relatively easily seen via
the di-lepton signal.\Ref\dilepton{F. Anselmo \etal, {\it Nuovo Cimento}
{\bf 106} (1993) 1389.}\ Nonetheless, sleptons provide more coverage,
since the $\cpone$ (and other ino's) are too heavy to
be detectable over about half of the allowed parameter space.
The lightest slepton, $\stauone$,
is likely to be detected at LEP-200 for $\mstauone\lsim 90-95\gev$. The
contour corresponding to 100 $\stauone^+\stauone^-$ events,
shown in Fig.~\one, indicates that detection will be possible
for all but a small section of parameter space if $\tau^+\tau^-+\ptmiss$
final states can be efficiently employed. $\slep^+\slep^-$ ($\ell=e,\mu$)
pairs have only slightly higher thresholds and would be easily detected in the
$\ell^+\ell^-+\ptmiss$ final state.  Since slepton-pair production
can indirectly probe $\mcpone$ values as large as $170\gev$, depending
on $\tanb$, it provides a deeper probe of the model
than chargino production.
\REF\LNPWZ{J. L. Lopez, D. V. Nanopoulos, H. Pois, X. Wang and A. Zichichi,
\prdj{48} (1993) 4062.} This is also discussed in Ref.~[\LNPWZ].

What about the Higgs bosons? First, we note that
$\mhl<110 (105)\gev$ for $\mu>0 (\mu<0)$.
Fig.~\one\ shows the contour for 100
$Z\hl$ events (the internal solid line) at LEP-200
(before cuts or branching ratios), as being an upper limit for $\hl$ discovery.
Note that $Z\hl$ events will be observable
for almost all of parameter space, in particular in that part of
parameter space where $\stauone$ detection becomes questionable.
Because $\mha\gsim 185\gev$ for the allowed parameter region (for
both signs of $\mu$), none of the other Higgs bosons will be detectable
at LEP-200. As already noted (see Fig.~\one) invisible decays of the $\hl$ will
be important, and can even be dominant, at smaller $\mha$ values.
However, one can still obtain
a Higgs signature via mass reconstruction using the recoiling $Z$ in $Z\hl$
events. Fig.~\one\ shows that the combined
$\hl$ and $\slep^\pm$ searches should completely explore the
YUMS parameter space for $\mu>0$ except for the
small window outlined by the $\hl Z$ and $\stauone^+\stauone^-$
contours in the vicinity of $3.5\lsim \tanb\lsim 4.5$,
$450\gev \lsim \mha \lsim 500\gev$, where no new particle could be detected.
For the $\mu<0$ case, there is no such window, and complete exploration
of the parameter space is possible at LEP-200. Even for $\mu>0$,
if $\Delta\rho<0.01$ is imposed there will be no window, $\hl$ discovery
being guaranteed.

The strongly interacting $\gl$ and $\sq\,$'s are most appropriately sought
at a hadron collider.  Since $\msq<\mgl$ throughout all of parameter
space, the $\sq\,$'s should provide the most direct signal. The decays
$\sq\rta q^\prime \cpmone,q\cnone$ are always allowed, with
(as noted above) $\cpone\rta \ell^+\snu+\slep^+\nu$,
as opposed to the more standard three-body modes. Generally, $\mcpone-\msnu$
and $\mslep-\mcnone$ (which determine the lepton spectra
in these two respective modes
\Ref\likesignnew{R.M. Barnett, J.F. Gunion and H. Haber,
\plbj{315} (1993) 349.})
while not large, are adequate to yield a reasonably hard $\ell$.
\REF\bghold{R.M. Barnett, J.F. Gunion and H. Haber, \prlj{60} (1988) 401.}
\REF\baermultilep{H. Baer, X. Tata, J. Woodside, \prdj{45} (1992) 142.}
Thus, generally speaking, isolated energetic leptons
can provide an excellent probe
for both the squarks and the gluino. (For the latter, the like-sign dilepton
signature can have a high rate.\refmark{\bghold,\baermultilep,\likesignnew})
However, the YUMS $\gl$ and $\sq$
mass scales are probably too large for discovery at the Tevatron.
In Fig.~\one, the contour for $\mgl=300\gev$ (dotted line)
represents our estimate of the absolute
maximum value that could possibly be probed by an upgraded Tevatron
(with $L\sim 1\fbi$),
even given the fact that the leptons for the like-sign dilepton
signature will (as noted above) be reasonably energetic.
Detection of the $\gl$ in the YUMS model seems unlikely
prior to the operation of the LHC. The lightest squark
is always the $\stop_1$ (the lightest of the two stops),
and we find $m_{\stop_1}\gsim 150\gev$.
\foot{Indeed, since the yukawa unification
condition forces $\mt$ to large values rather quickly for increasing
$\tanb$, there is even a small region where
$m_{\stop_1}<\mt$ (for $\tanb\lsim 1.6$, $\mt\simeq 160\gev$).}
The $\stop_1$ decays are fairly standard, with $\cpone b$ being
always allowed, while $\cnone t$ is allowed at higher $\mha$ values.
The other squarks are even heavier ($\msq\gsim
0.9 \mgl>250\gev$). These mass values are sufficiently large
that the Tevatron will also have considerable difficulty in discovering
the squarks until very high luminosity is available,
despite the favorable energetic lepton signature for the $\stop_1$ and
the other, heavier, $\sq\,$'s.\Ref\baertatagun{See, for example,
H. Baer, M. Drees, R. Godbole, J.F. Gunion and X. Tata,
\prdj{44} (1991) 725.}\
Of course, the inos and sleptons are much lighter; however,
they have lower production rates. Existing studies of ino detection
at a high luminosity Tevatron\Ref\charginotevatron{H. Baer and X. Tata,
\prdj{47} (1993) 2739; H. Baer, C. Kao and X Tata, \prdj{48} (1993) 5175;
J.L. Lopez, D.V. Nanopoulos, X. Wang and A. Zichichi, \prdj{48} (1993) 2062;
R. Barbieri \etal, \npbj{367} (1991) 28.}
suggest that extracting a tri-lepton signal (from $\cpone\rta \ell^+\snu,
\slep^+\nu$ and $\cntwo\rta \slep\ell$, with $\slep \rta \ell\cnone$)
should certainly be possible for $\mcpone\lsim 100\gev$, possibly extending to
$\mcpone\lsim 165\gev$ (see the dotdash curve in Fig.~\one).

\FIG\two{Contours for Higgs detection at the $4\sigma$ level at the LHC
for $L=100\fbi$: (left internal solid line) $\hl$ detection in
the $\ell\gam\gam$ mode;
(dotdash) $\hl$ detection in the $\gam\gam$ inclusive mode;
(dashes) $\hl$ detection in the $t\anti t b\anti b$ mode; (dotdotdash)
$\hl$ detection in the inclusive invisible
$\hl\rta(\cnone\cnone+\snu\snubar)$
mode. In the first three cases, the viable regions are above and to the right
of the contours; in the fourth case, the viable region is below and to
the left. We also show the contour for $\mgl=600\gev$ (dots), and
(right solid internal line) the contour for $m_{\tilde t_1}=400\gev$.
The line with long dashes shows the $B_0=0$ constraint between $\tanb$ and
$\mha$.  For this graph we have taken $\mu>0$.}

\topinsert
\vbox{\phantom{0}\vskip 5.0in
\phantom{0}
\vskip .5in
\hskip -20pt
\special{ insert user$1:[jfgucd.pois]noscale_lhc.ps}
\vskip -1.45in }
\centerline{\vbox{\hsize=12.4cm
\Tenpoint
\baselineskip=12pt
\noindent
Figure~\two:
Contours for Higgs detection at the $4\sigma$ level at the LHC
for $L=100\fbi$: (left internal solid line) $\hl$ detection in
the $\ell\gam\gam$ mode;
(dotdash) $\hl$ detection in the $\gam\gam$ inclusive mode;
(dashes) $\hl$ detection in the $t\anti t b\anti b$ mode; (dotdotdash)
$\hl$ detection in the inclusive invisible
$\hl\rta(\cnone\cnone+\snu\snubar)$
mode. In the first three cases, the viable regions are above and to the right
of the contours; in the fourth case, the viable region is below and to
the left. We also show the contour for $\mgl=600\gev$ (dots), and
(right solid internal line) the contour for $m_{\tilde t_1}=400\gev$.
The line with long dashes shows the $B_0=0$ constraint between $\tanb$ and
$\mha$.  For this graph we have taken $\mu>0$.
}}
\endinsert

At the LHC, there should be no difficulty in detecting the strongly
interacting gluino and squarks of the YUMS.  For instance,
gluinos heavier than $700\gev$ are not allowed,
and, as discussed, these will be easily discovered in either
the missing energy or like-sign dilepton modes.
(See Ref.~[\sparticles] and Refs.~[\bghold,\baermultilep,\likesignnew].)
Similarly, squarks much heavier than $\sim 540\gev$
are not allowed by the general model constraints, and multi-lepton
signals should provide an excellent probe for $\sq\sq$ and $\sq\gl$
production. (See Refs.~[\sparticles,\baermultilep].) Direct slepton and
ino detection will be more difficult, but may prove possible.
\REF\aguila{F. del Aguila and L. Ametller, \plbj{261} (1991) 326.}
\REF\BAERetal{H. Baer, C.-h. Chen, F. Paige and X. Tata,
preprint FSU-HEP-931104 (1994).}
Direct slepton production at the LHC has been considered
in Refs.~[\aguila] and [\BAERetal]. Both conclude that $\mslep\lsim200-250\gev$
should be explored via the di-lepton signal within one LHC year.
However, the generic decay chain masses assumed in these studies differ
in detail from the predictions of the YUMS, and Ref.~[\BAERetal] assumes
that $\snu$'s are visible through $\snu\rta \ell^\mp\cpmone$ decays,
a mode not generally present in the YUMS model.
$ep$ collisions at the LHC would provide a cleaner environment
for slepton searches.\Ref\bartl{A. Bartl \etal, \zpcj{52} (1991) 677.}
Detection of ino pair production at the LHC was studied in
Refs.~[\sparticles,\charginotevatron], which focused on 3 and 5 lepton
final states. The conclusion was that
for generically chosen masses and decay chains
pair production of inos (which had to include
at least one of the heavier inos) should be detectable for
an ino spectrum characterized by modest $M$ and $\mu$
values (up to $\sim 200\gev$). A reanalysis of all the above results
in the YUMS context would be quite useful, given the fact that
the lighter inos can have two-body decays to leptons, thereby typically
yielding more energetic leptons, that would more easily pass appropriate cuts.

Turning to Higgs boson detection at the LHC, we display in
Fig.~\two\ the contours
for $\hl$ detection in the $\gamma\gamma,\ell\gamma\gamma,t\bar t b\bar b$
and invisible $(\hl\rta\cnone\cnone,\snu\snubar)$
modes;
\foot{Recall that although the $\snu$ is not the LSP, it is essentially
invisible for the bulk of parameter space where its
dominant decay channel will be $\snu\rta\cnone\nu$.}
each is viable for some region of the allowed parameter space.
We see that detection of the $\hl$ will be possible over all of parameter
space in at least one mode (most often several modes).
However, at low $\mha$ ($\mha\sim 200-240\gev$) the $\hl$ can
only be detected at the LHC via the invisible decay procedures.

Since $\tanb\lsim 10$ and $\mha\gsim 185\gev$,
detection of the $\hh,\ha,\hpm$ in SM particle final states
will be very difficult at the LHC.
The proposed modes, such as $\hh\rta 4\ell$, $\ha\rta\gam\gam$,
$\ha\rta Z\hl\rta Z\tau\tau$ for small to moderate $\tanb$, and
$\hh,\ha\rta\tau^+\tau^-$ at large $\tanb$, all fail
due to the fact that the $\hh,\ha\rta \wt\chi\wt\chi,\slep\slep$
decay modes have large branching ratios (see earlier
discussion) and therefore deplete significantly
the SM final state branching ratios assumed in the detector studies.
In our analysis, the possibility of detecting the $\hh$ and $\ha$
in the invisible $\cnone\cnone,\snu\snubar$ channels was examined;
$4\sigma$ signals are not achievable in the allowed parameter space.
Assessment of the prospects for heavy Higgs detection in visible
$\wt\chi\wt\chi$ and $\slep\slep$ modes at a hadron
supercollider has only just begun, but some optimism
seems warranted.\refmark\baersparticle\

For $\mu<0$, since $\mhl<105\gev$, $\mgl\lsim 500\gev$ (and therefore
$m_{\slep_R}\lsim 75\gev$), $\slep^+\slep^-$ and $\hl Z$ will be observed
at LEP-200 for the whole of the allowed parameter space. At the LHC,
the no-lose theorem for $\hl$ Higgs detection
still holds; however due to the
positive shift in $\mcnone$ for $\mu<0$, the kinematically allowed region
for $\hl\rta\cnone\cnone$ disappears.
Instead, the $b\anti b$ mode covers the small-$\tanb$, small-$\mha$ corner
of allowed parameter space, while for $\tanb\gsim 6, \mha\lsim 200\gev$
the $\hl\rta \snu\snubar$ decays are dominant and the invisible decay
detection procedures would be necessary.

\vskip .15in
\leftline{\bf 3. Conclusions}
\vskip .075in

We have determined the allowed parameter space and
explored the prospects for sparticle and Higgs detection
(at LEP-200, the Tevatron and the LHC) in the
yukawa-unified ($\lambda_b(M_U)=\lambda_{\tau}(M_U)$, $m_0=A=0$)
minimal supergravity model. The model is completely specified by the
low energy parameters $\tanb,\mha$ (and the sign of $\mu$).
The result of requiring yukawa unification in addition to the usual
demand that the LSP be neutral is the upper limit
$\tanb\lsim 10$. This is {\it much}
more restrictive than obtained in $m_0=A=0$ supergravity scenarios
that do not require yukawa unification.
The more restrictive $B_0=0$ model falls within the YUMS scenario,
yielding $\tanb\sim 4$, implying $\mt\sim 185\gev$ (for $\mb(\mb)=4.25\gev$).
Exploration of SUSY breaking scenarios that are more general
than the YUMS model and a thorough discussion of the bottom-up approach
will be presented in a future publication.\refmark\gplater\

With regard to experimental probes, we find that for $\mu<0$
the combined $\cpmone,\stauone^\pm,\slep^\pm,\hl$
searches at LEP-200 should completely explore the presently
allowed parameter space, yielding a definitive test of the model.
For $\mu>0$, Fig.~\one\ shows a small remaining window
which should be well-explored by the LHC.

We also emphasize that the 'no-lose'
theorem for $\hl$ Higgs detection at the LHC is maintained, although
the techniques developed for detecting the
invisible $\hl\rta \cnone\cnone,\snu\snubar$ decays could be crucial if
$\mha\lsim 275\gev$.
Although the $\hpm,\ha,\hh$ Higgs bosons are produced in significant
numbers at the LHC, the experimental and consistency constraints on
the parameter space imply that none of their SM decay modes will
yield viable signals.
However, their detection at the LHC may prove possible in the generally
dominant $\wt\chi\wt\chi,\slep\slep$ modes.
If not, the $\hl$ would be the {\it only} detectable Higgs boson
before construction of the NLC.  At that, a rather high energy NLC
may be required to see the $\hh,\ha,\hpm$.
Given that much of the allowed parameter space (at least for
$\mu>0$) lies in the $\mha\sim\mhh\sim\mhpm\gsim 250\gev$ domain for which
$\hh\ha$, $\hp\hm$ pair production are the only large cross section processes
in $\epem$ collisions,
it is clear that $\sqrt s=500\gev$ would be far from certain to allow
us to see these heavier scalars.
If the NLC is run in the $\gam\gam$ collider mode, for which
single $\ha,\hh$ production is possible, the reach could be
extended somewhat.\Ref\gh{J.F. Gunion and H.E. Haber, Proceedings
of {\it Research directions for the Decade}, Snowmass (1990), p. 469;
\prdj{48} (1993) 5109.}
At Fermilab, the gluino and squarks of the YUMS models are essentially beyond
detection for $L\lsim 1\fbi$ (with the possible exception of the $\stop_1$),
whereas the inos might be detectable. All should be
within easy reach of the LHC given that $\mgl\lsim 700(500)\gev$,
$\msq\lsim 630(450)\gev$ for $\mu>0(\mu<0)$ and the inos are
relatively light.
A linear $\epem$ collider with sufficient energy would
also allow discovery of a large selection of the super partners.

\smallskip
\centerline{\bf Acknowledgments}
\smallskip
This work has been supported in part by Department of Energy
grant \#DE-FG03-91ER40674
and by Texas National Research Laboratory grant \#RGFY93-330.

\refout
\bye